\documentclass{rspublic}
 
\usepackage{epsf}
\usepackage{longtable}
\usepackage{natbib}
\usepackage{epstopdf}
\usepackage{graphicx}

\begin{document} 
 
\title[Introduction]{Recent Advances in Coronal Heating: Introduction} 
 
\author[I.~De Moortel, P.~Browning]{Ineke De Moortel$^1$, Philippa Browning$^{2}$} 
 
\affiliation{$^1$School of Mathematics \& Statistics, University of St Andrews, North Haugh, St Andrews KY16 9SS, UK\\$^2$Jodrell Bank Centre for Astrophysics, University of Manchester, Manchester M13 9PL }

\maketitle 
 
\begin{abstract}{Sun; astrophysics; magnetic fields; corona; chromosphere; reconnection; MHD waves} 
The solar corona, the tenuous outer atmosphere of the Sun, is orders of magnitude hotter than the solar surface. This ``coronal heating problem'' requires the identification of a heat source to balance losses due to thermal conduction,  radiation and (in some locations) convection. The review papers in this Theme Issue present an overview of recent observational findings, large and small scale numerical modelling of physical processes occurring in the solar atmosphere and other aspects which may affect our understanding of the proposed heating mechanisms. At the same time, they also set out the directions and challenges which must be tackled by future research. In this brief introduction, we summarise some of the issues and  themes which re-occur throughout this volume.

\end{abstract} 
 
\section{The Coronal Heating Problem} 
The so-called ``coronal heating" problem has existed for 75 years, since the solar corona was first demonstrated to contain plasma with temperatures of 1 million degrees Kelvin and above, much higher than the photospheric surface temperature of around 6000 K (\citealt{Grotian1939,Edlen1942}). There have been many advances over this period, with great progress in the last decades through a series of space missions. Nevertheless, this remains one of the outstanding unsolved problems in astrophysics (e.g.~\citealt{Klimchuk2006, Parnell2012}). 

The solar corona is the tenuous outer atmosphere of the Sun, visible from Earth only during a total eclipse, but with predominant emission in EUV and X-ray wavelengths observable only from space. It is highly structured, containing many loops of different scales comprising closed magnetic fields as well as regions of open magnetic field. The coronal heating problem requires the identification of a heat source to balance losses due to thermal conduction,  radiation and (in some locations) convection. It is widely accepted  that coronal heating is associated with the magnetic field, which, through the Lorentz force, provides the dominant force in the low $\beta$ corona, and  it is clearly evident that the magnetic structure primarily determines the strong spatial inhomogeneity in coronal emission.

The standard view of coronal heating is that free magnetic energy is either built up in the corona or transported to the corona, due to shuffling of the magnetic field lines in the photosphere (e.g. \citealt{Browning1991}). The majority of coronal heating models proposed so far  invoke either dissipation of  magnetohydrodynamic (MHD) waves or magnetic reconnection. The relative importance of these two processes should be  determined by the  time-scale of the photospheric driver with respect to the Alfv\'en time in the corona. However, it is now clear that the picture is more complex than this simple classification would suggest.  The recent understanding is summarised in review articles (e.g.~\citealt{Klimchuk2006, Parnell2012}), with loop heating also reviewed by, for example, \cite {Reale2014}.

 Although there is still no consensus on which mechanism is responsible for heating the solar atmosphere, a few simple but relevant ``facts" are now supported by a strong body of observational evidence (see also the reviews by \cite{Klimchuk2015} and \cite{Schmelz2015} in this volume):
 \begin{itemize}
\item there is reconnection in the corona and magnetic reconnection can dissipate stored magnetic energy;
\item there are many wave modes in the corona (but their actual contribution to coronal heating is still unclear);
\item the solar corona cannot be treated in isolation but should be seen as part of the complex, coupled solar atmosphere (including the chromosphere);
\item coronal heating is intrinsically non-steady.
\end{itemize}

The complexity of the solar atmosphere and the wide range of both spatial and temporal scales on which physical processes are observed to occur mean that beyond the above list of fairly simple facts, much is either not known or not agreed (see e.g. Klimchuk's review in this volume). There is, however, increasing consensus that ``the" coronal heating mechanism is unlikely to exist, but rather coronal heating is  due to different mechanisms in different places and/or at different times. Furthermore, the traditional dichotomy between waves and reconnection is not appropriate, since there are interactions between these processes (waves may drive reconnection, and vice versa); also turbulence, which may combine aspects of both wave and reconnection scenarios, is likely to play an important role (see \cite{Velli2015} in this volume). Recent work proposing coronal heating through Alfv\'enic turbulence \citep{vanballegooijen2011} has attracted much attention.

It is useful to keep in mind that the solar atmosphere does not only have to be heated but has to be formed in the first place. In other words, the coronal heating problem does not only involve an energy cycle but also a mass cycle. In addition, energy is required to accelerate the solar wind. The way forward lies in combining all aspects of solar physics research: a thorough understanding of the observations (including the underlying atomic physics), both large and small scale numerical simulations and microphysics which goes beyond MHD. The challenge for the Solar Physics community lies in combining these different aspects in a way that allows us to make progress in understanding the complex, dynamic solar atmosphere and, ultimately, its effect on Earth -  as well as beginning to understand the nature and variety of hot coronae in other stars as discussed by Testa in this volume \citep{Testa2015}.

The review papers in this Theme Issue were presented at a Theo Murphy meeting held at Chicheley Hall (25-27 Aug 2014). They present a combination of recent observational findings, large and small scale numerical modelling of physical processes occurring in the solar atmosphere and other aspects which may affect our understanding of the proposed heating mechanisms. As an introduction, we summarise some of the issues and  themes which re-occurred throughout the meeting.

\section{Advances in observations relevant to  coronal heating}

Present understanding of the constraints and input provided by the observations is summarised by \cite{Schmelz2015}. There have been substantial advances in recent years, especially due to data with high spatial resolution at a range of wavelengths from the Solar Dynamic Observatory (SDO); also very high spatial resolution data from the HiC rocket flight, X-ray imaging from XRT/Hinode, spectroscopy from EIS/Hinode, and most recently, IRIS ({\it Interface Region Imaging Spectrometer}; \citealt{DePontieu2014}) looking at lower layers of the atmosphere.  Interpretation of data from such instruments is supported by advances in atomic physics. 

\subsection{Advances in observations}
Observations show that the properties of coronal loops differ substantially from expectations of simple scaling laws, with loops having unexpectedly constant temperature distributions and being over-dense \citep{Schmelz2015}. The detection of very hot plasma (T $>$ 5 MK) in active regions, as predicted from nanoflare heating models by \cite{Cargill1994}, is very significant, but requires further study with new instruments. Studies of the properties of loops, such as their widths, also provides valuable constraints on heating mechanisms \citep{Klimchuk2015}.

In recent years, there has been a wealth of observations of various waves and oscillations in the corona \citep{Arregui2015}. As  well as potentially being directly involved in heating, the detection of waves and oscillations has opened up the field of coronal seismology, potentially allowing parameters such as magnetic field strengths to be inferred (see e.g. \citealt{IDM2012} for a recent review).

Another breakthrough in recent observations is the availability of vector magnetograph data, allowing better reconstruction of force-free magnetic fields in the corona and some knowledge of their topology and free magnetic energy, which are an essential ingredient for the coronal heating process (\citealt{Parnell2015}). However, better determination of coronal magnetic fields remains as a challenge for future work.  Furthermore, improved knowledge of small-scale photospheric velocity fields, which are the drivers of coronal heating, is essential.

\subsection{Observational tests of coronal heating models}

A number of approaches are available allowing observations to discriminate between theoretical heating models.  
Analysis of large datasets, for example of multiple coronal loops, is allowing detailed quantitative comparisons with predictions of theoretical heating models. One approach to observational tests of models is to use {\bf scaling laws}, investigating how heating rates vary with field strength and loop length \citep{Mandrini2000}. Analysis of {\bf Differential Emission Measure} (DEM) distributions is proving a powerful tool for potentially distinguishing between heating mechanisms \citep{Schmelz2015, Cargill2015}. The presence of {\bf non-thermal particles} could provide an important signature of energy dissipation mechanisms, often  being associated with magnetic reconnection, and recent results from IRIS \citep{Testa2014}  present exciting evidence, albeit indirect, of their creation in small-scale heating events. Detailed individual {\bf case studies}, as for example described by \cite{Longcope2015}, also provide a means of comparing theory and observation; in the mentioned example, observations of flux emergence in an Active Region demonstrate quantitatively that energy dissipated by magnetic reconnection matches radiated power.

\section{Modelling Coronal Heating}

The idea that the corona may be heated by the combined effect of very many small transient heating events known as ``nanoflares" \citep{Parker1988} has been seminal. It should here be noted that these transient energy releases are often considered to involve dissipation by reconnection, as in large-scale flares, but the nanoflare scenario may apply for any transient heating mechanism. Indeed wave heating is also naturally transient; for example, due to the nonlinear coupling between plasma heating and damping. In fact,  all heating mechanisms so far proposed will give impulsive heating to some extent \citep{Klimchuk2015}, but one of the fundamental questions to be resolved is the ``degree of unsteadiness" \citep{Cargill2015}. Comparison of predictions from modelling with recent observations from missions such as SDO is beginning to shed light on this \citep{Cargill2015, Schmelz2015}.

Current sheets with energy dissipation by magnetic reconnection may be created in many different ways, including emergence of new flux \citep{Longcope2015}. Detailed models of the build up of energy in magnetic fields and the subsequent energy release include models of field braiding (as reviewed by \citealt{Wilmot-Smith2015}) and  relaxation following an ideal MHD instability in a twisted field (see \citealt{Bareford2015}). The process by which reconnection actually dissipates the energy requires further investigation - dissipation need not take place mainly through Ohmic resistivity within current sheets, but rather  in larger-scale structures such as shocks \citep{Longcope2015, Bareford2015}.

Heating due to the dissipation of waves has always been an integral part of the coronal heating debate, with wave heating mechanisms being particularly attractive in the open field corona. However, as recent observations have revealed that significant wave power is present in (closed) coronal loops, wave heating mechanisms are regaining the attention of the solar physics community (\citealt{Arregui2015}). An important point to note here is that (observed) wave damping does not automatically imply dissipation, and hence heating, as the two processes could potentially operate on very different spatial and temporal scales. Theoretical models face the challenge to confirm that the observed waves and oscillations do indeed contribute to coronal heating  on relevant timescales  (\citealt{Arregui2015, Parnell2012}).

\subsection{Increasing Complexity}

During the meeting, it was widely stated in discussions that ``more complex" models are needed. Whilst this is self-evidently true, it is also the case that complexity is not necessarily a benefit in itself, and it is a challenge to determine in what ways simple models need to be extended in order to provide genuine new insights. One example is the complex topology and inherently 3D nature of the coronal magnetic field, which has important implications for storage and dissipation of magnetic energy \citep{Parnell2015}. Such complex fields naturally provide many sites for magnetic reconnection. However, little is known about how wave damping mechanisms, derived usually in very idealised 1D or 2D field models, work in more complex fields. 

A common theme across many of the presentations in the meeting was the importance of multi-thread structures within loops. Careful analysis of observational data provides evidence of such structure \citep{Schmelz2015}, whose existence has implications for heating both through waves \citep{Arregui2015} and reconnection \citep{Longcope2015} and with obvious relevance for flux braiding experiments, instabilities and nanoflare modelling, amongst others. 

Crucial in the ``complexity'' debate is the availability of rapidly increasing computational power, allowing us to run ever more complex (magnetic) geometries, larger scale numerical simulations and include additional physics. Such simulations are immensely  valuable  in understanding the complex dynamics of the solar atmosphere on full active-region scales, as reviewed by \cite{Peter2015}. Many of these models incorporate ``forward modelling'' (synthesising observed emission), facilitating both qualitative and quantitative comparisons with observations.  However, it is important that numerical simulations are interpreted carefully as issues such as boundary conditions and numerical dissipation are likely to affect the outcome of the modelling. Large-scale simulations need to be complemented by other models as again, increasing simulations in size is not necessarily a benefit in itself.

\subsection{Beyond MHD}

Traditionally, coronal heating modelling has been undertaken within the MHD framework, which indeed provides a very good description for large-scale coronal phenomena. However, any effective dissipation mechanism must involve processes on much smaller scales, at which the fluid description breaks down. Indeed, current sheet widths predicted within  standard MHD reconnection models (around 1 m) are far smaller than the particle mean-free-path in the corona (about 50 km), and hence the fluid approximation is not valid at the dissipation scales. This means that the reconnection process is collisionless. Similarly, wave dissipation models such as resonant absorption invoke dissipation at  small spatial scales, where the MHD model fails. Furthermore, in the chromosphere, partial ionisation can also be significant (\citealt{Martinez-Sykora2015}). Thus, modelling frameworks beyond single-fluid MHD, including fully kinetic plasma models,  are evidently required to understand at least some aspects of coronal heating.

One of the biggest challenges for modelling is the vast range of length scales involved. The global scale of coronal structures is of the order of Mm, whilst the ion skin depth is about 1 - 10 m, the ion gyro-radius is 0.1 - 1 m, and the electron gyro-radius even smaller. At the present - and in the foreseeable future - no single numerical simulation  could cover such a wide range of scales. 

As mentioned above, the presence and properties on non-thermal particles is an important signature of the dissipation processes, but modelling their generation requires kinetic models.

\subsection{The Chromosphere}

The layer between the corona and the solar surface (photosphere) is called the chromosphere. Although not quite as hot as the corona ($T_{\text{chromosphere}} \sim 10^3-10^4$ K), the chromosphere is arguably harder to heat than the corona due to its increased mass, requiring a heating rate an order of magnitude larger than the required coronal heating rate. Indeed, it might even be said that coronal heating is only a side effect of chromospheric heating, and hence there is not really a coronal heating problem at all. 

IRIS observations are revolutionising our understanding of the ''interface'' region, i.e.~the chromosphere and the transition region. For example, \cite{DePontieu2014b} report on sub-arcsec scale twisting and torsional motions, associated with transition region temperatures throughout the lower solar atmosphere. \cite{Hansteen2014} find evidence for a plethora of rapidly varying ($\sim$ minutes) short, low-lying loops at transition-region temperatures, finally resolving the postulated ``unresolved fine structure'' (necessary to reconcile observed emission and velocities with models of the solar atmosphere). The classical, layered view of the solar atmosphere is further challenged by \cite{Peter2014} who find $10^5$K plasma, heated by reconnection, low down, near the solar surface and sandwiched between cooler layers. \cite{Tian2014} report on small-scale jets undergoing rapid heating to transition-region temperatures which could form an intermittent but persistent contribution to the solar wind. The complex nature of the lower atmosphere is also highlighted in simulations, such as described by \cite{Martinez-Sykora2015}, which reveal some of the processes involved in energy transfer within partially-ionised regions. Clearly, IRIS observations are confirming that the coupling between the chromosphere and the corona cannot be ignored, i.e.~that the chromosphere cannot merely be seen as a passive layer between the solar surface and the solar corona.

\section{Summary} 
Although by no means a complete overview, the combined articles in this Theme Issue of {\it Philosophical Transactions of the Royal Society A} provide a thorough insight into contemporary thinking on coronal heating. They demonstrate how new observations from instruments with high spatial and temporal resolution, combined with advanced numerical simulations and underpinned by developments in fundamental theory, have progressed our understanding of the complex physical processes involved in solar coronal heating. The discussions in the meeting and the papers in this volume also highlight what is not yet known about  coronal heating, and set out the directions and challenges which must be tackled by future research. Future space instruments including NuSTAR, MaGIXS and Solar-C are expected to provide data which will resolve some of the outstanding questions but this can only be achieved if modellers provide testable predictions, which can be confronted with data. 


\begin{acknowledgements}
The authors would like to thank the Royal Society for their support throughout the organisation of the Theo Murphy Meeting and the production of this volume as well as all their colleagues who attended the meeting and contributed to the discussions.
\end{acknowledgements} 
 
\bibliographystyle{rspublicnat}

\bibliography{references}

\end{document}